\documentclass[onecolumn,aps,amsmath,showpacs,amssymb]{revtex4}

\usepackage[latin1]{inputenc}
\usepackage{graphicx}
\usepackage{subfigure}
\usepackage{color}
\usepackage{dcolumn}
\usepackage{bm}

\newcommand{\be}{\begin{equation}} 
\newcommand{\ee}{\end{equation}}
\newcommand{\bea}{\begin{eqnarray}} 
\newcommand{\eea}{\end{eqnarray}}
\newcommand{\Tr}{\textrm{Tr}}

\begin{document}

\title{Deconfinement vs. chiral symmetry and higher representation matter}
\author{T.~K\"ah\"ar\"a$^{a,b}$}
\email{topi.kahara@jyu.fi}
\author{M. Ruggieri$^c$}
\email{marco.ruggieri@lns.infn.it}
\author{K.~Tuominen$^{a,b}$}
\email{kimmo.i.tuominen@phys.jyu.fi}
\affiliation{$^a$Department of Physics, University of Jyv\"askyl\"a, P.O.Box 35, FIN-40014 Jyv\"askyl\"a, Finland}
\affiliation{$^b$Helsinki Institute of Physics, P.O.Box 64, FIN-00014 University of Helsinki, Finland}
\affiliation{$^c$Department of Physics and Astronomy, University of Catania, Via S. Sofia 64, I-95125 Catania}

\begin{abstract}
The interplay of deconfinement and chiral symmetry restoration are considered in terms of effective theories. We generalize the earlier model studies by considering fermions in higher representations, and study the finite temperature phase diagrams of SU(2) and SU(3) gauge theories with two fermion flavors in fundamental, adjoint or two-index symmetric representations. We discuss our results in relation to recent lattice simulations on these theories and outline possible applications in the context of dynamical electroweak symmetry breaking. 
\end{abstract}

\maketitle

\section{Introduction}

Our understanding of low energy QCD, and properties of hot and dense matter and phase diagrams of strongly interacting theories in 
general, is rooted in exact and approximate symmetries. For QCD with light quarks, the spontaneous breaking of the chiral symmetry is 
manifested in the spectrum of hadronic states, while for pure gauge theory the properties of the phase transition between confinement 
and hot gluonic matter are captured by universality arguments based on the center symmetry. The perturbative analyses, although
developed to impressive orders \cite{Kajantie:2002wa}, are inapplicable in the vicinity of the phase transition. Since this is the most 
interesting region for applications, like the heavy ion collisions, for hot QCD, alternative methods must be applied. The first principle 
method is provided by the lattice simulations and complementary methods are provided by the gauge/gravity dualities and 
effective models.

During recent years, a simple model framework able to account for the observed behavior in QCD has been developed. The model 
studies have mostly concentrated on hot and dense QCD, but also the dependence on the symmetry breaking parameters has been 
investigated. In this paper we will consider the Nambu-Jona-Lasinio model~\cite{Nambu:1961tp,Nambu:1961fr}
(NJL in the following, see~\cite{Klevansky:1992qe,Buballa:2003qv,Vogl:1991qt} for reviews), enlarged with a coupling to the Polyakov 
loop~\cite{Meisinger:1995ih,Fukushima:2003fw,Roessner:2006xn} (PNJL model,
see~\cite{Fukushima:2008wg,Megias:2006bn,Megias:2006ke,Sasaki:2006ww,Ghosh:2007wy,
Ciminale:2007sr,Hell:2009by,Abuki:2008nm,Sakai:2008py,Sakai:2008um,Sakai:2009dv,
Sakai:2010rp,Costa:2009ae,Ruivo:2011fg,Ohnishi:2011jv,Blaschke:2011yv} and references therein for previous studies), 
and extend it by considering higher representation matter fields.  

Generally, in a gauge theory, the matter fields in other than the adjoint representation break the center symmetry explicitly. In order to understand generic 
properties of strongly interacting gauge theories with matter in arbitrary representation, one needs to understand the fate of the deconfining phase transition 
as a function of fermion representations. For example, for SU(3) gauge theory with two Dirac flavors in the adjoint representation there are two transitions 
clearly separated: the deconfinement, measured by a rapid rise in the average value of the Polyakov loop, is observed at temperatures ca. an order of 
magnitude smaller than the onset of chiral symmetry restoration \cite{Karsch:1998qj}. In comparison, for QCD with two flavors in the fundamental 
representation, only a single crossover with simultaneous rise of the Polyakov loop and decrease of the chiral condensate is observed. 

The main phenomenological motivation for this study is in the possible applications of such theories in 
the model building for electroweak symmetry breaking via strong dynamics.
As is well known, when coupled with the electroweak currents of the Standard Model, the chiral condensate of light QCD quarks breaks electroweak 
symmetry, and provides dynamically the masses for the electroweak gauge bosons. This scenario is, however, phenomenologically 
unrealistic: the masses of the $W$ and $Z$ bosons would be only $\sim 30$ GeV and, moreover, the triplet of pseudoscalar Goldstone 
boson appears in the physical spectrum rather than being absorbed into the longitudinal modes of the massive electroweak gauge 
bosons. Nevertheless, given the hierarchy and naturality problems of the elementary scalars, one can entertain the thought that in place 
of the scalar sector of the Standard Model, a dynamical mechanism of the type described above occurs in a new strongly interacting 
sector of the Standard Model; this is the vintage Technicolor introduced by Susskind and Weinberg in late '70s \cite{Susskind:1978ms,Weinberg:1979bn}. 
The original models, 
which simply replicate QCD-like dynamics, are mostly ruled out by the electroweak precision data \cite{Peskin:1991sw}. Currently the model building efforts 
are concentrated on models which are quasi-conformal, meaning that the dynamics of the theory is governed by  a quasi-stable infrared 
fixed point. Over a large hierarchy of scales the Technicolor coupling constant, then, evolves very slowly, walks, in contrast to the running 
of the precociously asymptotically free QCD coupling. Hence these theories are categorized as walking Technicolor. The perturbative 
beta-function gives a guide how to construct such theories: the competition between anti-screening non-abelian gauge fields and 
screening matter fields implies a critical number of flavors $N_c^\ast$ above which there exists a nontrivial infrared fixed point. 
Comparing the value of the fixed point implied by the two loop beta function and the critical coupling for the onset of the chiral symmetry 
breaking defines the lower boundary of the conformal window of the gauge theory under consideration. For walking Technicolor one 
wants to tune the matter content of the theory  to reside close to the lower boundary of the conformal window, but still remain in the 
confining phase. 

For fundamental representation fermions the conformal window is expected to lie, roughly, above $N_f\sim 4N_c$. If all these flavors are 
charged under the electroweak interactions in the usual manner, one typically encounters again large tension with the data from the 
electroweak precision measurements. This is so, since the so-called $S$-parameter effectively counts the new degrees of freedom 
contributing to the electroweak symmetry breaking. A way to alleviate this tension was proposed in \cite{Sannino:2004qp}: Since the fermions 
in higher representations carry more charge, they screen more efficiently, and as a consequence walking can be achieved with fewer 
flavors.

In this paper we initiate a systematic study of higher representations within the PNJL model with the aim of establishing the phase structure, in particular the intertwinement of deconfinement and chiral symmetry restoration.
We consider the strong dynamics in isolation, i.e. we do not consider the coupling with the electroweak currents. Also, to better illustrate the fixing of model parameters between theories with different fermion representations, 
we use as a benchmark model the PNJL model tuned to QCD, which will also allow us to compare our results with respect to the ones currently available in literature \cite{Zhang:2010kn}. In this paper our aim is to show that theories with higher representation matter fields lead to novel patterns of how deconfinement and chiral symmetry restoration intertwine; we expect these patterns to remain intact if the theory is scaled to correspond with the strong dynamics at the electroweak scale.  Quantitative applications for particular Technicolor models, which would also require the inclusion of electroweak currents, we leave for future work.

\section{Model framework and parameter setup}

\subsection{Model setup}

In this paper we consider the Polyakov extended Nambu--Jona--Lasinio (PNJL) model. Based on our earlier experience, we expect that similar results will be obtained also in the Polyakov extended quark meson models (PQM). The PNJL model is defined by the Lagrangian
\begin{eqnarray}
{\mathcal{L}}={\mathcal{L}}_{\rm{chiral}}+{\mathcal{L}}_{\rm{Polyakov}}+{\mathcal{L}}_{\rm{interaction}}.
\end{eqnarray}
The chiral part of the Lagrangian corresponds to the usual NJL model.
The pure gauge dynamics are encoded in ${\mathcal{L}}_{\rm{Polyakov}}$, which is essentially a parametric fit to lattice data. Finally,
the interaction term features a connection between the chiral and pure gauge sectors. This is achieved by coupling the quarks to a static 
background gauge field; the derivation has been exposed in numerous earlier works, see e.g. \cite{Ratti:2005jh,Kahara:2008yg}. 
For the purposes of this paper it is sufficient to state the end result which is the grand potential of the form
\begin{equation}
\Omega = U_{\chi}+U_\ell+\Omega_{\bar{q}q}.
\label{genpot}
\end{equation}

The chiral part, $U_{\chi}$, is given by
\begin{eqnarray}
U_{\chi} &=& \frac{(m_0-M)^2}{2G}, 
\label{chipots}
\end{eqnarray}
where constituent quark mass is $M=m_0-G\langle\bar{q}q\rangle$.

The gauge sector, $U_\ell$, is independent of the chiral model but depends on the chosen gauge
group. In this work we consider $SU(2)$ and $SU(3)$ gauge groups. The respective $Z(2)$ and $Z(3)$ symmetric 
forms for the pure gauge potential are
\begin{eqnarray}
\label{gaugepotsu2}
U_{\ell_{\rm{SU(2)}}}/T^4 &=&  -\frac{a(T)}{2}|\ell_{F_{SU(2)}}|^2 + b(T)\ln[1 - |\ell_{F_{SU(2)}}|^2] \\
U_{\ell_{\rm{SU(3)}}}/T^4 &=&  -\frac{a(T)}{2}|\ell_{F_{SU(3)}}|^2 + b(T)\ln[1 - 6|\ell_{F_{SU(3)}}|^2
 + 4(\ell_{F_{SU(3)}}^3 + \ell_{F_{SU(3)}}^{\ast3}) - 3(|\ell_{F_{SU(3)}}|^2)^2],
\label{gaugepotsu3}
\end{eqnarray}
where $\ell_F$ is the Polyakov loop in the fundamental representation of the gauge group.
The temperature dependent coefficients $a(T)$ and $b(T)$ are
\begin{equation}
a(T) = a_0 + a_1\left(\frac{T_0}{T}\right) + a_2\left(\frac{T_0}{T}\right)^2 \quad \textrm{and} \quad b(T) = b_3\left(\frac{T_0}{T}\right)^3, 
\end{equation}
where the parameters $a_0,a_1,a_2,b_3$ and $T_0$ will be fitted to pure gauge lattice data. The arguments of the logarithms
are based on the Haar--measure of the gauge group~\cite{Ratti:2005jh} and results in a divergence at $\ell_F = 1$, thus restricting 
the Polyakov loop to values $\ell_F < 1$. In what follows, we will assume that the Polyakov loop expectation value
is homogeneous in space. This assumption will simplify the evaluation of the fermion determinant; see below. 
The final part, $\Omega_{\bar{q}q}$, of the grand potential, contains the interaction terms between the chiral and gauge sectors.
It arises from a simple integration over the quark fields coupled to a background gauge field~\cite{Fukushima:2003fw}:
\begin{eqnarray}
\label{omegaqqbar1}
\Omega_{\bar{q}q} &=& -\text{dim}(R) 2N_f\int\frac{d^3p}{(2\pi)^3}E \nonumber \\
&&-2N_fT\int\frac{d^3p}{(2\pi)^3}\left({\rm{Tr}}_c\ln\left[1+L_Re^{-(E-\mu)/T}\right]+{\rm{Tr}}_c\ln\left[1
+L_R^\dagger e^{-(E+\mu)/T}\right]\right),
\end{eqnarray}
where $E = \sqrt{\vec{p}^2 + M^2}$ and the constituent mass $M$ is given
below equation (\ref{chipots}). Finite baryon chemical potential, $\mu$, will not be considered in this
paper, and in subsequent equations we assume that $\mu = 0$. In the above equation, 
the dimension of the fermion representation, $\text{dim}(R)$, is made explicit.  In the above equation the Wilson line in the representation $R$ 
of the gauge group is defined as
\begin{eqnarray}
L_R(\vec{x}) &=& P\exp\left[-\int_0^{1/T} d\tau A^a_0(\vec{x},\tau)T^a(R)\right]\nonumber \\
&=& \exp\left[-A_0^aT^a(R)/T\right],
\end{eqnarray}
where the latter equality holds for a static background field. The Polyakov loop is then defined as the trace 
\begin{equation}
\ell_R(\vec{x})=\frac{1}{{\rm{dim}}(R)}{\rm{Tr}}L_R(\vec{x}).
\label{pldef}
\end{equation}

The first term on the r.h.s. of Eq.~\eqref{omegaqqbar1}, which represents the
vacuum fluctuations, is divergent and needs to be regulated.  In PQM, the renormalizability of the model allows for a rigorous 
renormalization procedure of the divergence~\cite{Skokov:2010sf}.
On the other hand, in the case of the PNJL model we will use a momentum dependent NJL coupling, 
\begin{equation}
G(p) = G \theta(\Lambda - |p|)~.
\label{eq:PPP}
\end{equation}
In the vacuum term, this amounts to cut off the condensate dependent integral at the scale $\Lambda$,
thus making the latter convergent (the divergence appears in a condensate independent integral which
can be subtracted). The thermal part of (\ref{omegaqqbar1}) is convergent,
hence it does not need regularization. However, the prescription~\eqref{eq:PPP} implies
that for $p>\Lambda$ the free gas contribution $M = m_0$ has to be taken. 
These modes do not give an explicit contribution to the chiral condensate; however,
they are $Z_3$ charged and thus couple to the Polyakov loop. As a consequence, they
contribute to the thermodynamics of the system and have to be taken into account.
The prescription in ~\eqref{eq:PPP} corresponds to an oversimplified version of
a nonlocal and static NJL vertex. In our study, the choice~\eqref{eq:PPP} is justified a posteriori,
since only in this case we obtain clearly separated deconfinement and chiral restoration
in the case of adjoint fermions. This result is in agreement with what has been found
in~\cite{Zhang:2010kn} where two colors QCD with adjoint fermions has been considered. 
As a matter of fact, in~\cite{Zhang:2010kn} a cutoff on the thermal part of the free energy density is introduced,
in order to have a net separation of the two QCD transitions in agreement with the lattice results. 
 
The thermodynamics of the system is obtained by
solving the equations of motion for the order parameters of the system
\begin{eqnarray}
\frac{\partial\Omega}{\partial M}=0, ~~\frac{\partial\Omega}{\partial\langle\ell\rangle}=0,
\label{eom}
\end{eqnarray}
and then evaluating the grand potential at the minimum to obtain the free energy at given temperature.
We work under the mean field approximation, replacing the Polyakov loops with their thermal expectation values, $\ell \rightarrow \langle\ell\rangle$. 

\subsection{Introducing higher representation fermions}
Our main objective is
the study of fermions in higher representations. Besides these, we consider fermions in the fundamental
representations of the color group in order to fix the model parameters. For the SU(2)
gauge group we use the fundamental and adjoint representations while for the standard SU(3) case we include also the
two--index symmetric representation, i.e. the sextet, in our study. 
As explained in the Introduction, these cases are relevant for the phenomenology of walking technicolor theories.
In this section we will
describe how we construct the fundamental and higher Polyakov loops and give the form of the interaction potential (\ref{omegaqqbar1})
for the different representations.

In the fundamental representation of a $SU(N)$ gauge group, the Wilson line $L_F$ can be written as a diagonal
matrix
\begin{equation}
L_F = \textrm{diag}(e^{i\theta_1}, \cdots, e^{i\theta_N}),
\end{equation}
with the constraint $\theta_N = -(\theta_1 + \cdots + \theta_{N-1})$. For $SU(2)$ this means that there is only one free
parameter $\theta$; for the $SU(3)$ group there are two independent parameters, $\theta_1$ and $\theta_2$. 
For simplicity, in the $SU(3)$ case we choose $\theta_2 = 0$ which leaves us with one free parameter 
$\theta_1 = \theta$. This assumption is justified at zero quark chemical
potential, which is the case we consider in this article, since the Polyakov loop is real, $\ell = \ell^\dagger$.

The Wilson line for the adjoint representation (for both $SU(2)$ and $SU(3)$) can be constructed from the fundamental
Wilson lines using the relation
\begin{equation}
L_A^{ab} = 2\Tr[L_F T^a L_F^{\dagger} T^b],
\end{equation}
where $T^i$'s are the generators normalized as $\Tr[T^a T^b]=\delta^{ab}/2$. In the $SU(3)$ case, in addition to the adjoint Wilson line, we want
to know the Wilson line in the two index symmetric representation. The Wilson line for this representation is related
to the fundamental ones through a similar relation
\begin{equation}
L_6^{ab} = \Tr[L_F T_6^a L_F T_6^b],
\label{sextetloop}
\end{equation}
where the set of matrices matrices $T_6^i$ $(i = 1 \cdots 6)$ is a basis for symmetric $3 \times 3$--matrices and given explicitly in Appendix \ref{symm}. 

The Polyakov loops
can be calculated as a function of $\theta$ according to the definition (\ref{pldef}). For the fundamental representation one
gets
\begin{eqnarray}
\ell_{F_{SU(2)}} &=& \cos \theta \\
\ell_{F_{SU(3)}} &=& \frac{1}{3}(1 + 2 \cos \theta).
\end{eqnarray}
The higher representation loops can be similarly calculated and expressed in terms of the fundamental loops
\begin{eqnarray}
\ell_{A_{SU(2)}} &=& \frac{1}{3}(2\cos2\theta + 1) = \frac{1}{3}(4\ell_{F_{SU(2)}}^2 - 1) \label{fsu2ar}\\
\ell_{A_{SU(3)}} &=& \frac{1}{8}(2 + 4\cos\theta + 2\cos2\theta) = \frac{1}{8}(9\ell_{F_{SU(3)}}^2 - 1) \label{fsu3ar}\\
\ell_{6_{SU(3)}} &=& \frac{1}{6}(2 + 2\cos\theta + 2\cos2\theta) = \frac{1}{6}(9\ell_{F_{SU(3)}}^2 - 3\ell_{F_{SU(3)}}) \label{fsu3sr}.
\end{eqnarray}

One can now proceed to perform the color traces in potential (\ref{omegaqqbar1}) for the different gauge groups
and representations. For this we write the interaction part of the potential as
\begin{equation}
\label{omegaqqbar2}
\Omega_{\bar{q}q} = 
-4N_fT\int\frac{d^3p}{(2\pi)^3}\left(S_R \right)-\text{dim}(R)2 N_f\int\frac{d^3p}{(2\pi)^3}E,
\end{equation}
where we have defined
\begin{equation}
S_R \equiv \Tr_c \ln[1 + L_R e^{-E/T}].
\end{equation}
Performing the traces in SU(2) we find 
\begin{equation}
S_{F_{SU(2)}} = \ln[1 + 2\ell_{F_{SU(2)}}e^{-E/T} + e^{-2E/T}]
\end{equation}
and
\begin{equation}
S_{A_{SU(2)}} = \ln[1 + e^{-E/T}] + \ln[1 + (4\ell_{F_{SU(2)}} - 2)e^{-E/T} + e^{-2E/T}].
\end{equation}
Analogously, in the case of SU(3) we find
\begin{equation}
S_{F_{SU(3)}} = \ln[1 + e^{-E/T}] + \ln[1 + (3\ell_{F_{SU(3)}} - 1)e^{-E/T} + e^{-2E/T}],
\end{equation}
\begin{equation}
\begin{aligned}
S_{A_{SU(3)}} &= 2\ln[1 + e^{-E/T}] + 2\ln[1 + (3\ell_{F_{SU(3)}} - 1)e^{-E/T} + e^{-2E/T}] \\
&+ \ln[1 + (9\ell_{F_{SU(3)}}^2 - 6\ell_{F_{SU(3)}} - 1)e^{-E/T} + e^{-2E/T}],
\end{aligned}
\end{equation}
and finally
\begin{equation}
\begin{aligned}
S_{6_{SU(3)}} &= 2\ln[1 + e^{-E/T}] + \ln[1 + (9\ell_{F_{SU(3)}}^2 - 3\ell_{F_{SU(3)}} - 2)(e^{-E/T} + e^{-3E/T}) \\
&+ 3(1 + \ell_{F_{SU(3)}} - 9\ell_{F_{SU(3)}}^2 + 9\ell_{F_{SU(3)}}^3)e^{-2E/T} + e^{-4E/T}].
\end{aligned}
\end{equation}
We have written the color traces for the different representations in terms of the fundamental representation
loop of the corresponding gauge group. Proceeding this way we keep the same observables, 
the constituent mass and the fundamental Polyakov loop,
when comparing the different representations and their effects on the chiral and deconfinement transitions. 

\subsection{Fixing the parameters}

In this section we discuss how the model parameters, fixed for three colors and fundamental representation fermions  can be used to
obtain the parameters in the case of higher representation matter or two colors.
There are three parameters which enter into the chiral sector of the PNJL model. 
These are the bare quark mass $m_0$, the value of the four--fermion coupling $G$ 
and the momentum cut--off $\Lambda$. For the SU(3) fundamental fermions, 
we fix the parameters as usual~\cite{Asakawa:1989bq,Klevansky:1992qe,Vogl:1991qt,Hatsuda:1994pi} by requiring that
the physical values of $m_\pi$, $f_\pi$ and $\langle\bar{q}q\rangle$ are reproduced.
The parameter set we use is shown in the first row of Table \ref{modpar}.

\begin{table}
	\begin{center}
	\caption{Parameters of the NJL model and derived observable values for different cases. The SU(3) fundamental values
	 correspond to observed QCD properties, while the rest are obtained by scaling the coupling $G$ as explained in the text. }
	\begin{tabular}{c | c  c  c | c  c  c  c}
	\hline
	 & \multicolumn{3}{c |}{NJL parameters} & \multicolumn{4}{c}{Derived observable values} \\
	\hline
	 & $m_0$ & $G$ & $\Lambda$ & $f_\pi$ & $m_\pi$ & $m_\sigma$ & $M$ \\
	\hline
	\bf{SU(3) Fundamental}& 5 MeV & 8.879 GeV$^{-2}$  & 678 MeV & 93 MeV & 138 MeV & 599 MeV & 296 MeV \\
	\bf{SU(2) Fundamental}& 5 MeV & 13.32 GeV$^{-2}$ & 678 MeV & 76 MeV & 137 MeV & 598 MeV & 296 MeV \\
	\bf{SU(3) Adjoint}& 5 MeV & 7.492 GeV$^{-2}$  & 678 MeV & 167 MeV & 166 MeV & 2567 MeV & 1157 MeV \\
	\bf{SU(2) Adjoint}& 5 MeV & 23.66 GeV$^{-2}$ & 678 MeV & 96 MeV & 179 MeV & 2811 MeV & 1406 MeV \\
	\bf{SU(3) Sextet}& 5 MeV & 11.099 GeV$^{-2}$ & 678 MeV & 139 MeV & 174 MeV & 2615 MeV & 1304 MeV \\
	\hline
	\end{tabular}
	\label{modpar}
	\end{center}
\end{table}

In addition to the parameters of the chiral sector, the parameters for the pure gauge potentials  (\ref{gaugepotsu3}) have to be assigned. 
This can be done by fitting the potential to pure gauge lattice data. In this work we will use  for SU(3) the fit presented in
\cite{Roessner:2006xn} and the required coefficients are shown in Table \ref{gaugepar}.
\begin{table}
	\caption{Lattice data fitted coefficients for the SU(3) and SU(2) pure gauge potentials (\ref{gaugepotsu3}) and (\ref{gaugepotsu2}), respectively.
	Values for SU(3) are obtained from \cite{Roessner:2006xn}, while SU(2) values are fitted in this work to lattice data from \cite{Engels:1981qx}.}
	\begin{center}
	\begin{tabular}{ c  c  c  c  c}
	\hline
	\multicolumn{2}{c}{} & SU(3) & \multicolumn{2}{c}{}  \\
	\hline
	 $a_0$ & $a_1$ & $a_2$ & $b_3$ & $T_0$\\
	 3.51 & -2.47 & 15.2 & -1.75 & 270 MeV \\
	\hline 
	\multicolumn{2}{c}{} & SU(2) & \multicolumn{2}{c}{} \\
	\hline
	 $a_0$ & $a_1$ & $a_2$ & $b_3$ & $T_0$\\
	 1.19 & -1.136 & 7.94 & -2.759 & 268 MeV \\
	\hline 
	\end{tabular}
	\end{center}
	\label{gaugepar}
\end{table}

With the set of parameters in Tables~\ref{modpar} and~\ref{gaugepar}, the 
thermodynamic potential for the PNJL model in the case of SU(3) with fundamental quarks
is completely specified. 

To discuss higher representation fermions and different numbers of colors, we need a scheme to translate the model parameters 
between different cases. 
As a matter of fact, in the case of representations different from the fundamental one of SU(3), 
the numerical values of physical observables are not known. As a consequence, we need 
some argument to obtain the parameters for the higher representations, from those
fixed for the fundamental representation. We will first discuss how the parameters are obtained for higher representations
of SU(3) and then how to obtain the parameters for the two color theory.

Of the three parameters, $m_0$, $G$ and $\Lambda$, the quark mass is basically a parameter that determines the amount 
of explicit chiral symmetry breaking in the theory
and is thus responsible also for the mass of the pion. It is essentially a free parameter even in QCD so there is no need
to adjust it when changing fermion representations. The remaining parameters $G$ and $\Lambda$ are connected to the fundamental theory,
in the case under consideration to QCD with two colors and/or fermions in different representations.
For what concerns the cutoff, we do not change its numerical value
after changing the fermion representation. A more rigorous treatment
would require the fixing of $\Lambda$ (and of $G$ as well) by reproducing the phenomenological
properties of the vacuum as in the case of three colors discussed above. However, data
about these properties are missing. In absence of a serious guiding principle, 
we use the same value of $\Lambda$ for all the cases considered in this article.

On the other hand, we have a theoretical guiding principle to
change the value of $G$ according to the dimension of the representation considered,
which was introduced in~\cite{Zhang:2010kn}. It is 
based on the Fierz transformation properties of the color current interaction
\begin{equation}
\mathcal{L}_{\text{int}} = -g_s(\bar{\psi}\gamma^\mu T_{aR}\psi)^2
\end{equation}
in which the coupling $g_s$ can be related to
the underlying QCD coupling and the NJL coupling $G$ \cite{Buballa:2003qv}.
For an effective NJL interaction of fermions in representation $R$,
\begin{equation}
\mathcal{L}_{\text{NJLint}} = -G_R(\bar{q}q)^2~,
\end{equation}
one gets
\begin{equation}
G_R = g_s\frac{C_2(R)}{\text{dim}~R}~,
\label{coupscale}
\end{equation}
where $C_2(R)$ is the quadratic Casimir operator for representation $R$. For the derivation
of this result we refer to the original article~\cite{Zhang:2010kn}. 
Assuming that the coupling $g_s$ is the same for all representations, the ratio of the NJL
couplings for different representations will be independent of $g_s$:
\begin{equation}
\frac{G_{R_1}}{G_{R_2}} = \frac{C_2(R_1)}{C_2(R_2)}\frac{\text{dim}~R_2}{\text{dim}~R_1}~.
\end{equation}
By means of the above equation we are able to determine the higher representation NJL coupling 
from the fundamental one, once the Casimir operators for the
relevant representations of SU(N) are specified. These operators for the fundamental,
adjoint and symmetric representations are shown in Table~\ref{casimirs}. 
Together with the fundamental representation ones, the model parameters for higher fermion representations
are collected In Table \ref{modpar}.

\begin{table}
	\caption{Quadratic Casimir operators and representation dimensions for SU(N).}
	\begin{center}
	\begin{tabular}{ c | c  c }
	\hline
	 Representation & $C_2$ & dim \\
	\hline
	 Fundamental & $\frac{N^2-1}{2N}$ & $N$ \\
	 Adjoint & $N$ & $N^2-1$  \\
	 Symmetric & $\frac{(N-1)(N+2)}{N}$ & $\frac{N(N+1)}{2}$  \\
	\hline 
	\end{tabular}
	\end{center}
	\label{casimirs}
\end{table}

Then, let us consider setting up the model parameters in SU(2) case. As in the SU(3) case with higher representation matter,
we have no experimental data to fit that would determine our model parameters. Therefore we again make use of scaling arguments
to infer the SU(2) parameters from the known SU(3) ones. 
In particular, in the large $N_c$ limit of QCD (see~\cite{Jenkins:2009wm} for a review) one finds $f_\pi\propto N_c^{1/2}$ and
$\langle\bar{q}q\rangle \propto N_c$. These scalings are respected by the NJL model,
see for example~\cite{Klevansky:1992qe}.
Therefore it is natural to fix the values of $G$ and $\Lambda$, for fermions in the
fundamental representation of SU(2), by requiring that the values of the properly scaled $f_\pi$
and $\langle\bar{q}q\rangle$ are reproduced. For what concerns the value of the bare quark mass,
the GMOR relation $f_\pi^2 m_\pi^2 \propto m_0 \langle\bar{q}q\rangle$ together with the scaling
$m_\pi \propto N_c^0$ implies $m_0 \propto N_c^0$, hence $m_0$ does not need to be changed. 

In the NJL setup used here, this is consistent with assuming that the four--fermion coupling $G$ is proportional to $1/N_c$. To see this,
one may first note that the NJL gap equation, with cutoff $\Lambda\sim N_c^0$, implies $M\sim N_c^0$. Then, since the gap equation
can be written as $M=m_0+2G\langle\bar{q}q\rangle$, one further obtains $\langle\bar{q}q\rangle\sim N_c$ consistently with the large $N_c$ expectation discussed above. For the determination of $f_\pi$ and $m_\pi$ are determined via Eqs. (4.26) and (4.21) of \cite{Klevansky:1992qe}, respectively, and these imply that $f_\pi\sim N_c$ and $m_\pi\sim N_c^0$ again consistently with the expected large $N_c$ behavior.

Thus going from fundamental SU(3) to fundamental SU(2) we multiply $G$ by $3/2$ and, to keep things
as simple as possible, we again do not alter the momentum cutoff. As a result of this scaling choice
the pion and sigma vacuum masses, calculated from the NJL model, remain the same between fundamental SU(3) and fundamental SU(2).
Then, to  define the model corresponding to SU(2) with two adjoint fermions we use the same Fierz scaling argument that we utilized
in the SU(3) case. The resulting parameter values for both fundamental and adjoint SU(2) are also shown in Table  \ref{modpar}.

\begin{figure}[htb]
\centering
  \subfigure{
  \hskip-1.0truecm
  \includegraphics[width=8.5cm]{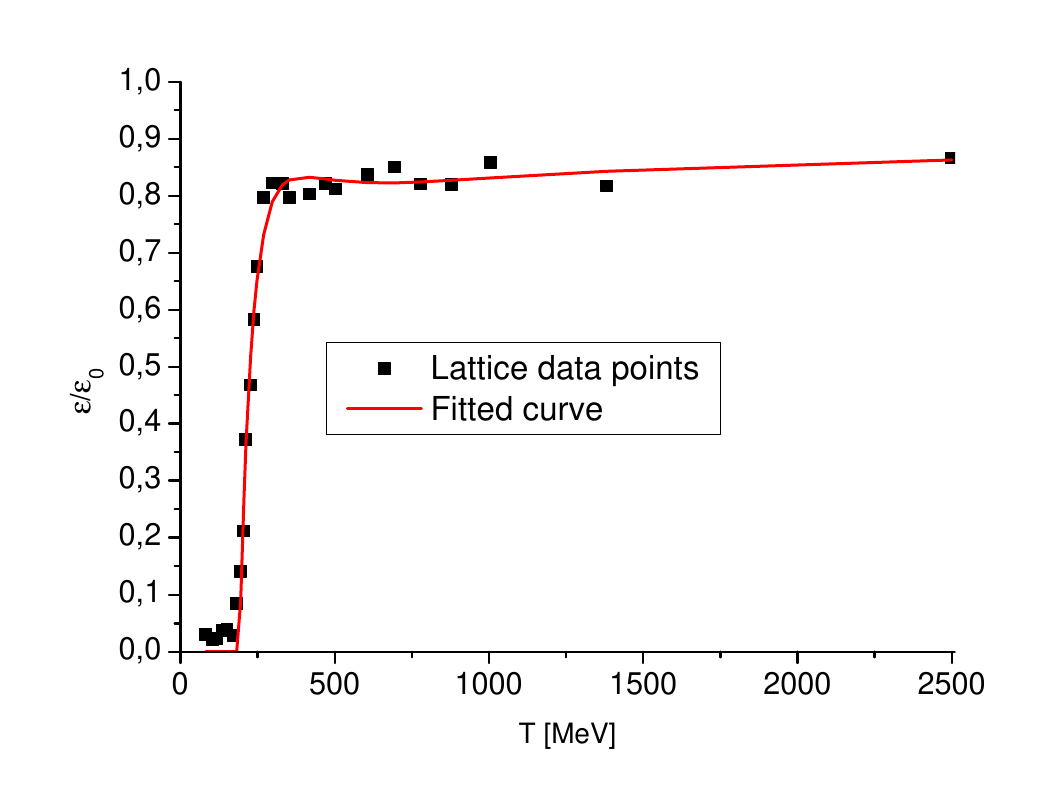}}
\vskip-0.4truecm
\caption{The data points show the energy density of SU(2) pure gauge theory~\cite{Engels:1981qx}, 
while the solid curve shows the corresponding result obtained from the Polyakov loop potential \eqref{gaugepotsu2} with parameters as given in Table \ref{gaugepar}.}
\label{FITG}
\end{figure}

Finally, since also the pure gauge sector is changed when switching from SU(3) to SU(2), new parameters have to be assigned for the pure gauge potential
(\ref{gaugepotsu2}) as well. As in the SU(3) case the coefficients are fitted to reproduce lattice data. We use SU(2) pure gauge results from 
~\cite{Engels:1994xj} and fit
the coefficients of the potential in Eq.~\eqref{gaugepotsu2} so that it 
reproduces a matching energy density as a function of temperature with the data.
The data with the resulting fit are shown in Figure~\ref{FITG}, 
while the numerical values of the coefficients $a_0,a_1,a_2,b_3$ and $T_0$ are listed
in Table \ref{gaugepar}.

\section{Numerical results}

In this section we establish the finite temperature phase diagrams of PNJL model with 
fermions in fundamental and higher representations of the SU(2) and SU(3) gauge groups by
numerically evaluating the temperature dependence of the Polyakov loop condensate and 
the constituent mass. These are the (approximate) order parameters for the chiral and deconfinement transitions, 
and their temperature dependence displays the interplay between the deconfinement and the chiral
restoration crossovers.

First, we report on our results on the SU(2) gauge group. This case has been
already studied in the literature by means of effective models coupled to the
Polyakov loop~\cite{Brauner:2009gu,Zhang:2010kn}; therefore the results presented
here have to be considered as a check of our calculations. Nevertheless,
they are quite important since they are useful to interprete the analogous results
obtained in the case of the SU(3) gauge group. Before going ahead, we notice that in the numerical computations with fermions,
we do not change the value of $T_0$ in the Polyakov loop effective potential.
This is done just for a matter of simplicity; in principle,
an explicit dependence of $T_0$ on the number of active flavors should be
introduced \cite{Schaefer:2007pw}.
This choice for simplicity, however, allows us to focus only on the 
few parameters in the chiral sector, which will be changed according to the representation
and to the gauge group as explained in the previous subsections 
(at the end of the day, we just will modify the coupling constant). This is helpful to
realize what is the main source of the different behavior of deconfinement
and chiral symmetry restoration when different representations for fermions
are considered. 

Finally, we ignore the explicit dependence of the NJL coupling on the Polyakov loop,
as studied in the recent literature~\cite{Kondo:2010ts,Sakai:2010rp,Sakai:2011gs,Gatto:2010pt}.
Also this choice is inspired by simplicity. Together with other compromises we have made in fixing
the values of the model parameters, also this aspect has
a quantitative impact on the interplay between deconfinement and chiral symmetry restoration,
and certainly it deserves further study which we leave for a future project.

\begin{figure}[htb]
\centering
  \subfigure{
  \hskip-1.0truecm
  \includegraphics[width=8.5cm]{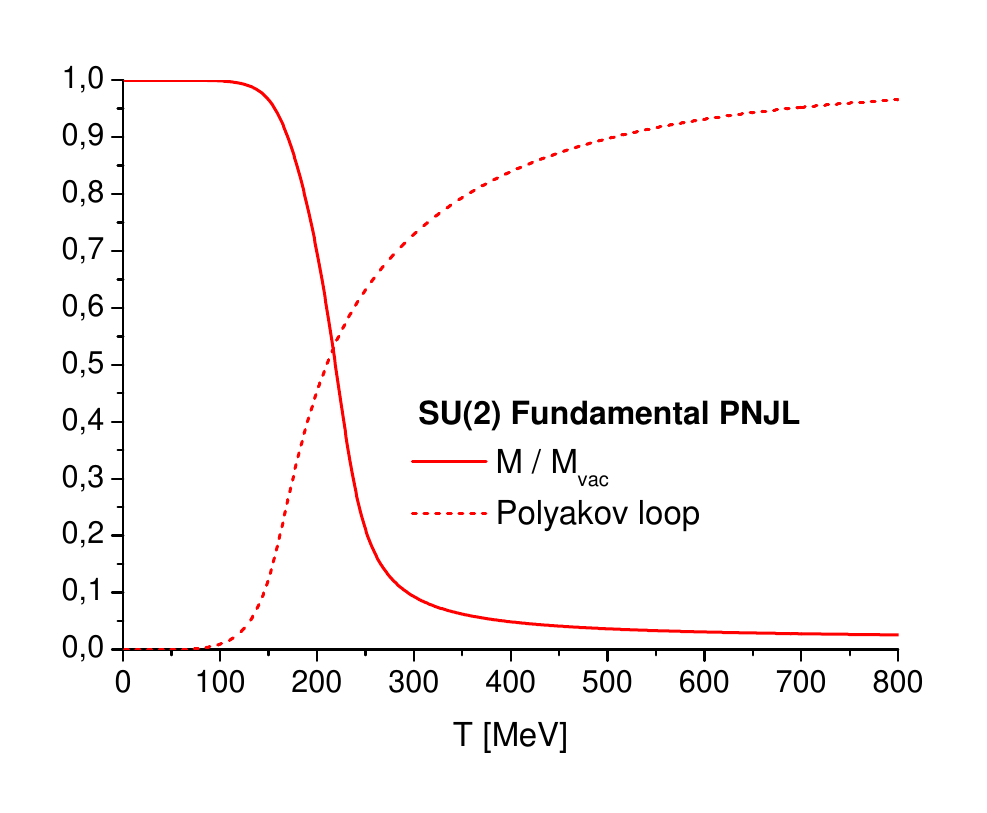}}
\vskip-0.4truecm
\caption{The chiral and deconfinement order parameters as a function of temperature for the PNJL model 
in the SU(2) fundamental fermion case. Model parameters listed in Table \ref{modpar}.}
\label{SU2FOPG}
\end{figure}

\begin{figure}[htb]
\centering
  \subfigure{
  \hskip-1.0truecm
  \includegraphics[width=8.5cm]{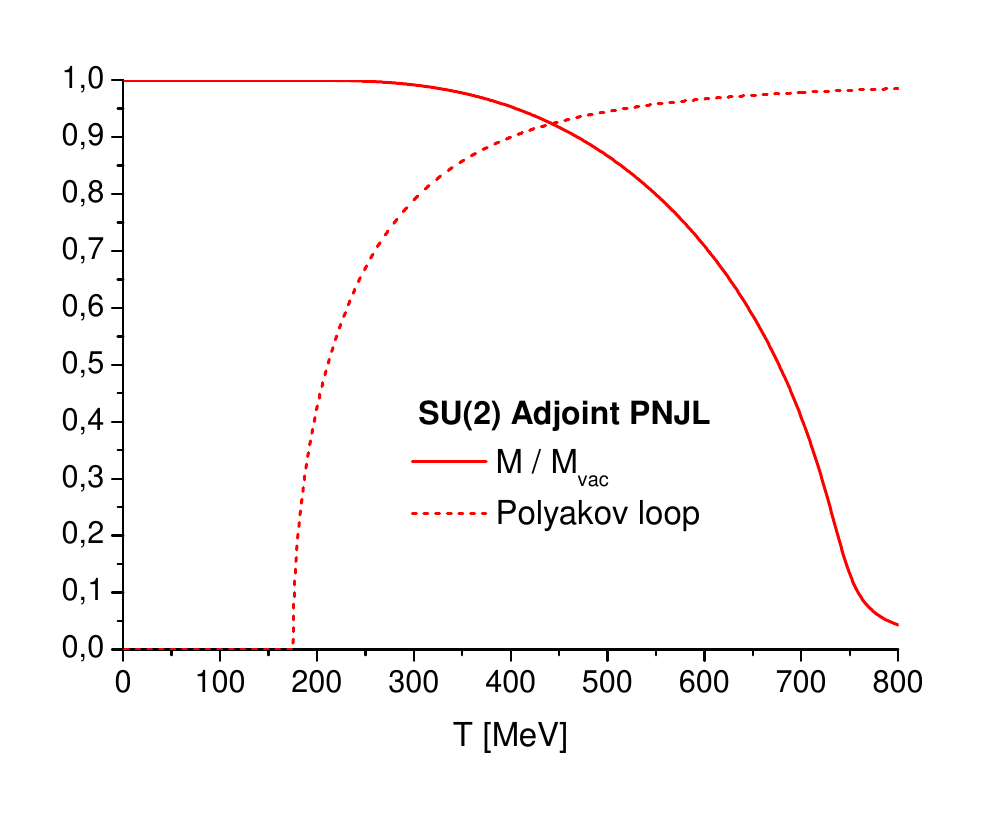}}
\vskip-0.4truecm
\caption{The chiral and deconfinement order parameters as a function of temperature for the PNJL model 
in the SU(2) adjoint fermion case. Model parameters listed in Table \ref{modpar}.}
\label{SU2AOPG}
\end{figure}

\begin{figure}[htb]
\centering
  \subfigure{
  \hskip-1.0truecm
  \includegraphics[width=8.5cm]{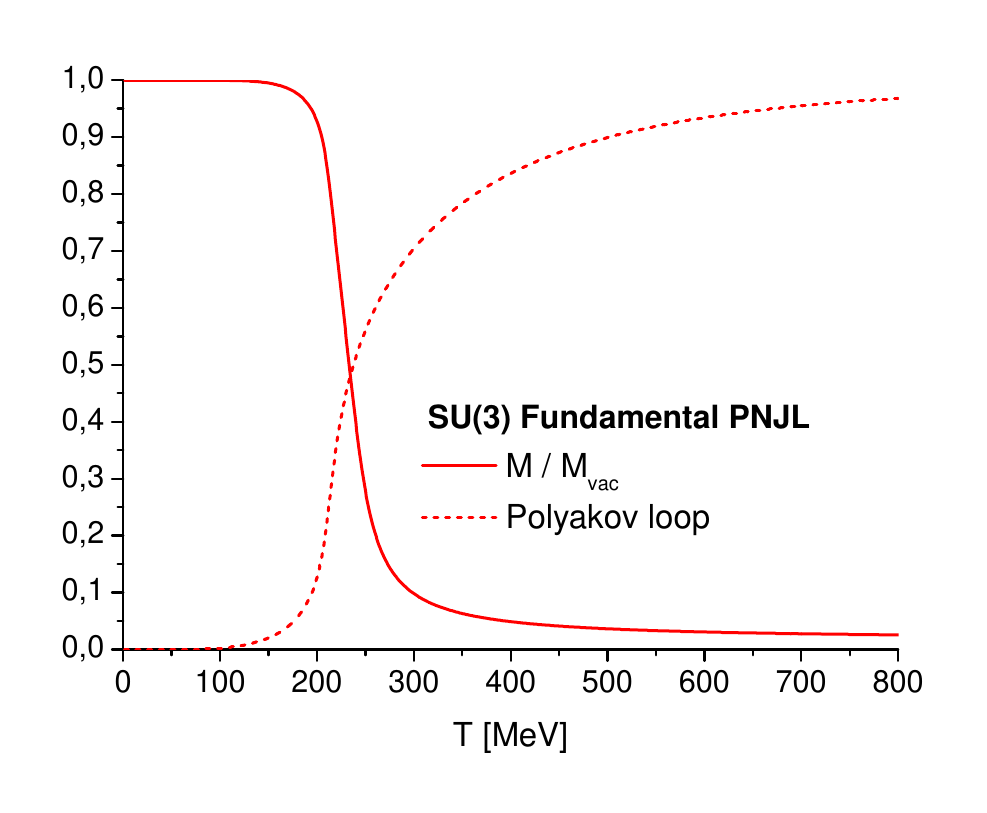}}
\vskip-0.4truecm
\caption{The chiral and deconfinement order parameters as a function of temperature for the PNJL model 
in the SU(3) fundamental fermion case. Model parameters listed in Table \ref{modpar}.}
\label{SU3FOPG}
\end{figure}

For the SU(2) case, the chiral and deconfinement
order parameters as a function of temperature are shown in Figures \ref{SU2FOPG} and \ref{SU2AOPG} for
the fundamental and adjoint fermions, respectively. In the case of fundamental fermions 
we expect the behavior of the order parameters be similar with the
more familiar SU(3) fundamental fermion case; indeed both chiral and
deconfinement transitions are smooth cross--overs which are located quite close together. However, for a crossover
transition the exact transition temperature depends largely on the definition one uses. From Figure \ref{SU2FOPG}
one could say that both transitions happen near $T \sim 220$ MeV since near this temperature both the normalised
constituent mass and the Polyakov loop cross the value $0.5$. If one makes a comparison with the respective SU(3)
case, shown in Figure \ref{SU3FOPG}, one can note that perhaps the only significant difference
is the slightly reduced transition temperature and a smoother rise in the value of the Polyakov loop when approaching the phase transition 
region $T = 200 - 220$ MeV. This close similarity can also be expected by looking at Table \ref{modpar} where
the pion, sigma and constituent quark masses show no deviation between the SU(2) fundamental and SU(3) fundamental cases.
As explained earlier, this is due to the scaling of $G$ being partially cancelled by the change in the number of 
color degrees of freedom.

In the case with adjoint fermions the situation is different from the fundamental case. A large separation
between the two transitions appears, and the deconfinement transition has changed qualitatively. 
This is in agreement with earlier studies~\cite{Brauner:2009gu} and with 
the lattice results.
In more detail, the pseudocritical temperature of the chiral cross--over 
is much higher than the one found in the case of fundamental fermions, namely around $700$ MeV. 
On the other hand, 
the deconfinement transition is now a true second order phase transition near $T \sim 175$ MeV. 
The latter aspect is easily understood: Fundamental fermions break the center symmetry,
thus act as a source for the Polyakov loop, and Z(2) breaking persists in the confining phase,
turning the deconfinement transition to a crossover. On the other hand, adjoint fermions
do not break Z(2), which turns to be a true symmetry of the action, unbroken
in the confinement phase. This is what one would expect solely on the basis of 
underlying symmetries \cite{Mocsy:2003qw}.  

\begin{figure}[htb]
\centering
  \subfigure{
  \hskip-1.0truecm
  \includegraphics[width=8.5cm]{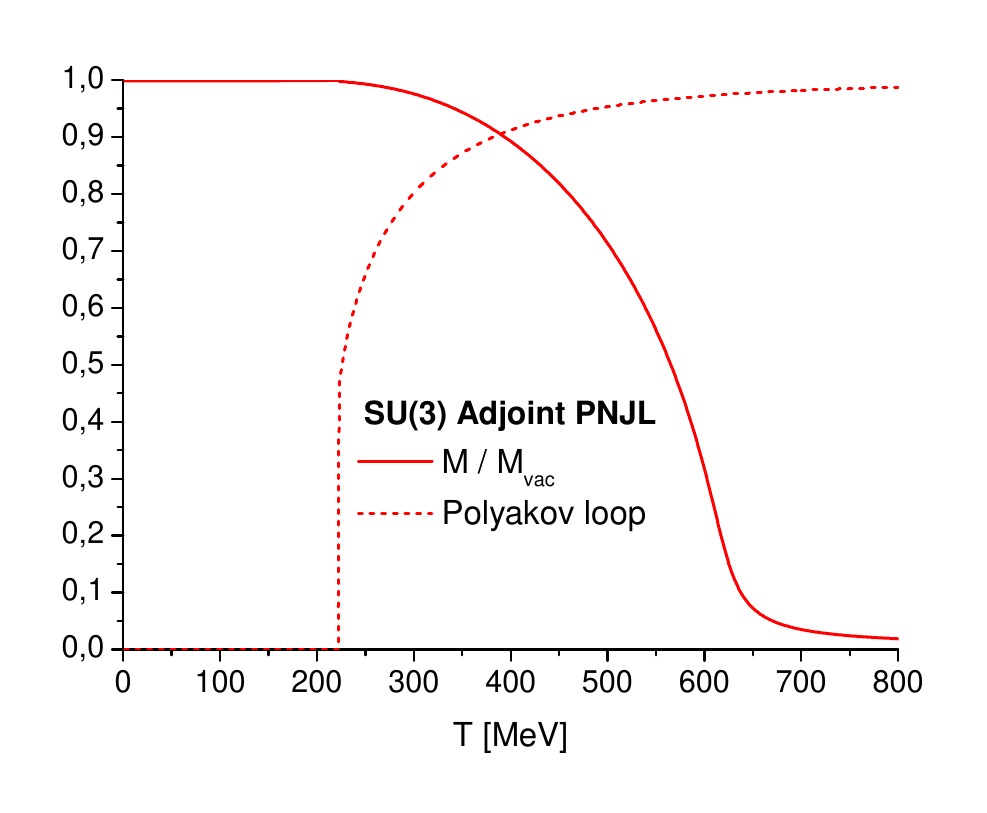}}
\vskip-0.4truecm
\caption{The chiral and deconfinement order parameters as a function of temperature for the PNJL model 
in the SU(3) adjoint fermion case. Model parameters listed in Table \ref{modpar}.}
\label{SU3AOPG}
\end{figure}

\begin{figure}[htb]
\centering
  \subfigure{
  \hskip-1.0truecm
  \includegraphics[width=8.5cm]{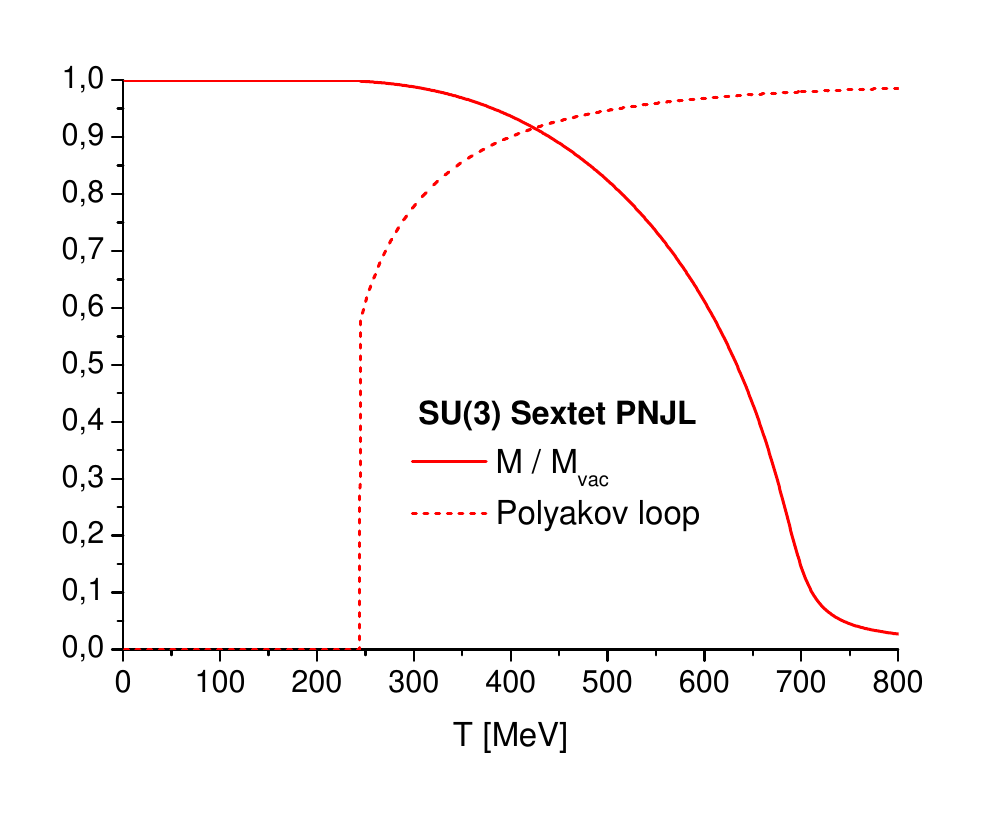}}
\vskip-0.4truecm
\caption{The chiral and deconfinement order parameters as a function of temperature for the PNJL model 
in the SU(3) sextet fermion case. Model parameters listed in Table \ref{modpar}.}
\label{SU3SOPG}
\end{figure}

Let us then turn to the numerical results on SU(3) theory with fermions in fundamental, adjoint or sextet representation.
As in the SU(2) case, we consider the behavior of the order parameters 
as a function of temperature.
The results are shown in Figures \ref{SU3FOPG}, \ref{SU3AOPG} and \ref{SU3SOPG} for the fundamental, adjoint and sextet
representations, respectively. 

The order parameters for 
the case of fundamental SU(3) fermions show only minor quantitative differences with the respective SU(2) case and
feature the typical coincidence of the two crossovers in a range of temperature
centered on $T_c \approx 230$ MeV.
Note that this crossover temperature is a little
higher than in the earlier works \cite{Kahara:2008yg,Kahara:2009sq,Kahara:2010wh}.
The difference is small, $\sim 5$ percent, and is not of importance for the results of this work. 
The discrepancy is due to two reasons. First, the dynamical mass cutoff in thermal contribution of quarks,
arising from Eq.~\eqref{eq:PPP}, was
not accommodated in the calculations of \cite{Kahara:2008yg,Kahara:2009sq,Kahara:2010wh}.
Second, we are using here 
slightly different initial parameter set than in~\cite{Kahara:2008yg,Kahara:2009sq,Kahara:2010wh}, but its
effect on the transition temperatures is very small. 

As in the SU(2) case, the temperature dependence of the order parameters for adjoint fermions of SU(3) correspond to
expectations based on the approximate symmetries of the order parameters. 
The NJL model exhibits separate transitions with roughly a $400$ MeV gap between the two. Again the
chiral transition is a smooth crossover while the Polyakov loop retains the first order transition from the pure
gauge sector. The deconfinement, which is a crossover for fundamental representation matter fields,  is turned to a true
phase transition because quarks in the adjoint representation do not break the center
symmetry Z(3) explicitly.

Finally, we turn to the more interesting case of the PNJL model with sextet fermions, which has not yet been studied 
in the chiral model literature.
The SU(3) sextet results are remarkably similar with the SU(3) adjoint case: both phase transition temperatures deviate less than
20 percent from the adjoint, shifted towards higher temperatures, and the qualitative features of the phase transition are nearly identical.
At a first sight, this behavior of the order parameters might appear unexpected, since the center
symmetry is explicitly broken by fermions in the sextet representation.

Our interpretation of the results is as follows:
In the case of sextet fermions, the constituent quark mass at zero temperature is considerably higher than
the mass of fermions in the fundamental representation. This is the natural consequence of having 
a larger NJL coupling in the sextet representation. When quarks couple to the Polyakov loop,
the amount of breaking of the center symmetry is proportional to the hopping parameter, which in turn
scales as $1/m$ where $m$ is the quark mass. In the model at hand, at the one-loop level
the mass eigenstates which propagate in the Polyakov loop background 
have mass $m=M$ with $M$ being the constituent quark mass. The latter is
very large in the case of the sextet fermions, which causes the hopping parameter to be very small
in this case. Thus, the explicit breaking of the center symmetry, even if present,
is very soft, and the Polyakov loop behaves effectively as if the center symmetry
was unbroken. 

\section{Conclusions}

In this paper we have initiated a study of PNJL model for applications beyond QCD. In particular we concentrated on the SU(2) and SU(3) gauge theories with two Dirac fermion flavors in higher representations of the gauge group, as these have been established as a phenomenologically viable theories for dynamical electroweak symmetry breaking. Eventhough these theories are interesting for understanding general features of strong dynamics, the real motivation for our study is that if such strong dynamics underlies the Higgs sector to be uncovered at the LHC, it becomes immediately of interest to study their finite temperature phase diagrams for the evolution of the early universe. Based on the experience gained in investigations of hot and dense QCD matter, the PNJL model provides a good quantitative tool for such studies. As a novel new result we have shown that, while the theories with fundamental and adjoint fermions show the behavior expected by considering the underlying symmetries, the SU(3) theory with two sextet fermions is, despite the explicit breaking of the center symmetry, akin to the theory with adjoint fermions where the center symmetry is not broken by the presence of matter fields. 

One can imagine several further avenues for research. First, all the theories with higher representation matter we have considered are currently investigated 
in lattice simulations, see e.g. \cite{Hietanen:2008mr,DelDebbio:2009fd,DeGrand:2008kx,Kogut:2010cz} . Similarly as with the case of QCD, for which the PNJL model has been successfully applied, for these theories we expect the future 
developments and increased precision in the lattice analyses of these theories to help in better fixing the model parameters. For applications to electroweak phase transition in the early universe one should couple the electroweak currents. For theories with only chiral fields see \cite{Jarvinen:2009wr} for recent developments. We aim to proceed towards these directions taking also into account the nontrivial behaviors due to the deconfinement transition which we derived in this paper.

\acknowledgements
The work of T.K. was supported by the EU-FP7-IA-Hadron Physics2 TORIC WP, n. 4007601. The work of M.R. is supported by the FIRB program "Phases of QCD, field and transport theory".

\appendix
\section{A basis for symmetric $3\times 3$ matrices}
\label{symm}

To define the Polyakov loop in the sextet representation, we use the following basis for symmetric $3\times 3$ matrices:
\bea
T_6^1 &=& \left( \begin{array}{ccc} 1 & 0 & 0 \\ 0 & 0 & 0 \\ 0 & 0 & 0 \end{array}\right)\qquad
T_6^2 = \left( \begin{array}{ccc} 0 & 1/\sqrt{2} & 0 \\ 1/\sqrt{2}  & 0 & 0 \\ 0 & 0 & 0 \end{array}\right) \nonumber \\
T_6^3 &=& \left( \begin{array}{ccc} 0 & 0 & 0 \\ 0 & 1 & 0 \\ 0 & 0 & 0 \end{array}\right)\qquad
T_6^4 = \left( \begin{array}{ccc} 0 & 0 & 0 \\ 0 & 0 & 1/\sqrt{2} \\ 0 & 1/\sqrt{2} & 0 \end{array}\right) \nonumber \\
T_6^5 &=& \left( \begin{array}{ccc} 0 & 0 & 0 \\ 0 & 0 & 0 \\ 0 & 0 & 1 \end{array}\right)\qquad
T_6^6 =\left( \begin{array}{ccc} 0 & 0 & 1/\sqrt{2} \\ 0 & 0 & 0 \\ 1/\sqrt{2} & 0 & 0 \end{array}\right).\nonumber
\eea

\bibliographystyle{ieeetr}
\bibliography{bibliography}

\end{document}